\documentclass[twocolumn,10pt]{IEEEtran}
\usepackage{algorithm,algorithmic,amsbsy,amsmath,amssymb,epsfig,bbm,mathrsfs,fancyhdr,fancyvrb,subfigure,url,tabularx,graphicx}
\usepackage[export]{adjustbox}
\usepackage{cases,comment,color,rotating,footnote,epstopdf,multirow,pifont,inputenc}
\usepackage[doublespacing]{setspace}

\newcommand{\beq}{\begin{equation}}
\newcommand{\eeq}{\end{equation}}
\newcommand{\beqn}{\begin{eqnarray}}
\newcommand{\eeqn}{\end{eqnarray}}

\newcommand{\qed}{\nobreak \ifvmode \relax \else
      \ifdim\lastskip<1.5em \hskip-\lastskip
      \hskip1.5em plus0em minus0.5em \fi \nobreak
      \vrule height0.75em width0.5em depth0.25em\fi}
			
			\DeclareMathOperator*{\argmax}{argmax}
      
\begin{document}

\title{Minority Games With Applications to Distributed Decision Making and Control in Wireless Networks}
\author{Shermila Ranadheera, Setareh Maghsudi, and Ekram Hossain}

\maketitle

\begin{abstract}
Fifth generation (5G) dense small cell networks (SCNs) are expected to meet the thousand-fold mobile traffic challenge within the next few years. When developing solution schemes for resource allocation problems in such networks, conventional centralized control is no longer viable due to excessive computational complexity and large signaling overhead caused by the large number of users and network nodes in such a network. Instead, distributed resource allocation (or decision making) methods with low complexity would be desirable to make the network self-organizing and autonomous. Minority game (MG) has recently gained attention of the research community as a tool to model and solve distributed resource allocation problems. The main objective of this article is to study the applicability of the MG to solve the distributed decision making problems in future wireless networks. We present the fundamental theoretical aspects of basic MG, some variants of MG, and the notion of equilibrium. We also study the current state-of-the-art on the applications of MGs in communication networks. Furthermore, we describe an example application of MG to SCNs, where the problem of computation offloading by users in an SCN is modeled and analyzed using MG.
\end{abstract}


\section{Introduction}
\label{intro}
The next generation of wireless networks, also known as 5G, is expected to face a thousand-fold growth in mobile data traffic due to the increased smart device usage, proliferation of data hungry applications and pervasive connectivity requirement. Since the existing traditional macro cellular networks are not designed to cope with such large data traffic, network densification using small cell base stations (SBS) and implementation of small cell networks (SCNs) are proposed. In particular, SCNs are expected to improve the efficiency of the utilization of radio resources, including energy and spectrum. 

Although SCNs might become the key enablers of 5G, they impose some challenges that need to be addressed. For instance, the typical wireless resource allocation problems become more complicated in a dense network. Since the SCNs are expected to be hyper-dense and multi-tier, they must be self-organizing and self-healing, to avoid high complexity and fault-intolerance of central management. In other words, network management tasks such as resource allocation are preferred to be performed in a distributed manner. Also the unavailability of global and precise channel state information in dense networks needs to be addressed.  Moreover, feedback and signaling overhead should be minimized. 
In order to address these challenges, in this paper we focus on the minority game (MG) and its potential applications to solve the distributed decision making/control problems that arise in 5G SCNs.

Minority game has recently gained attention of the research community as a tool to model congestion problems encountered in wireless networks. In simple terms, in an MG, an odd number of players select between two alternatives in the hope of being in the minority, because only the minority group receives a pay-off. Thus, an MG is able to model a congested system with a large number of agents competing for shared resources, where pair-wise communication between agents does not take place. This finds application in 5G SCNs that accommodate a large number of users, where congestion can occur due to the scarcity of the (radio and/or computational) resources. In such scenarios, users would naturally prefer to select the less-crowded option. Moreover, MG involves self-organized decision making with minimal external information available to the agents as desired in dense SCNs. 

When developing distributed solution schemes for wireless resource allocation problems, conventional distributed approaches (e.g. those based on traditional game theory) are not always applicable, since such models become increasingly complex for systems with large number of agents. In essence, many such game models require pairwise interactions among the agents. In contrast, the agents' interaction in MG exhibit mean field like behavior \cite{korutcheva2004advances}; i.e., an individual agent interacts with the aggregate behavior of all other agents. This makes MG a promising technique, especially since mean field based models are widely used as fitting tools to model large systems that are often studied in distributed resource allocation problems. 


The rest of the article is organized as follows. In Section II,  the basic concepts, some variants, and equilibrium notions of MG are discussed. Section III describes the state-of-the-art applications of minority games in communication networks and outlines potential future applications and research directions. In Section IV, an example application of MG for  distributed computational offloading is presented before the article is concluded in Section V.

\section{Minority Games: Basics, Equilibrium, Solution Approaches, and Variants}
\label{tutorial}

\subsection{Basics of a Minority Game}
The concept of MG stems from \textit{El Farol bar problem} \cite{RePEc:aea:aecrev:v:84:y:1994:i:2:p:406-11}, and was initially formulated and presented in \cite{RePEc:eee:phsmap:v:246:y:1997:i:3:p:407-418}.  In the most basic setting of such a game, an odd number of players choose between two actions while competing to be in the minority group through selecting the less popular action, since only the minority receives a reward. After each round of play, all players are informed of the winning action, which is then used as history data by the players to improve the decision making in the upcoming rounds. Let us denote the two actions by $0$ and $1$. Moreover, the action of player $i$ at time $t$ is shown by $a_i(t)$. The number of players ($N$) is required to be an odd number to avoid ties. Each player has a given set of decision making strategies that help her select future actions. A \textit{strategy} predicts the winning action of the next round based on the previous $m$ number of winning actions, with $m$ being the size of memory, also known as the \textit{brain size}. In other words, a \textit{strategy} is essentially a mapping of the $m$-bit length history string ($\mu(t)$) to an action. An example strategy table for an agent is given in Table \ref{table1}, where the agent has two strategies $S_1$ and $S_2$. Since there are two actions to select from, it is clear that the strategy space consists of $2^{2^m}$ total number of strategies, which become very large even for small $m$. Thus, \textit{reduced strategy space} (RSS) is introduced to make the strategy space remarkably smaller without any significant impact on the dynamics of the MG. RSS is formulated by choosing $2^m$ strategy pairs so that in each pair, one strategy is \textit{anti-correlated} to the other. In other words, the predictions given by one strategy are the exact opposites of the predictions given by the other strategy \cite{Yeung2009}. 
An example for two anti-correlated strategies is shown in Table \ref{table1}. Thus RSS constitutes of $2^{m+1}$ total number of strategies, which is much smaller than the size of universal strategy space $2^{2^m}$.  


\begin{table}[h!]
  \centering
  \caption{An example strategy table for an agent}
  \label{table1}
\begin{tabular}{|*{6}{c|}}  
\hline
History string & \multicolumn{2}{c|}{Predicted winning action} \\ \hline
 & \multicolumn{1}{c}{$ \quad \quad   S_1$} \hspace{0.8cm}& \multicolumn{1}{c|}{$S_2$}\\ \hline 
00 & 1 & 0 \\ \hline
01 & 1 & 0  \\ \hline
10 & 0 & 1 \\ \hline
11 & 0 & 1  \\ \hline
\end{tabular}
\end{table}

At the outset of the game, each agent randomly draws $S$ strategies from the strategy space which remain fixed for each player throughout the game. There is no a priori best strategy. Intuitively, if such strategy exists, all agents would use it and therefore lose due to the minority rule, which contradicts the initial assumption. As the game is played iteratively, each player evaluates her own strategies as follows: The strategies that make accurate predictions about the winning action are given a point and the poorly performing strategies are penalized. In other words, strategies are reinforced as they predict the winning action over a number of plays. Note that all strategies are scored after each round regardless of being used by the agent or not. Thus the score of each strategy is updated after each round of play according to its performance and the players use the strategy with the largest accumulated score at each round. Each player's objective is to maximize her utility over the time as she plays the game repeatedly. In MG, often the players compete for a limited resource without communicating with each other. Consequently, since players do not have any knowledge about other players' decisions, the decision making becomes almost autonomous \cite{korutcheva2004advances}. 

\subsection{Properties of a Minority Game}
\label{MGproperties}  

The properties of a minority game are described by the following parameters and behaviors:

\begin{itemize}
\item{\textit{Attendance}:} One of the most important parameters of an MG is the collective sum of the actions of all players at a given time $t$, known as the \textit{attendance}, $A(t)$. 
\item{\em Volatility}:
Basically, the attendance value never settles but fluctuates around the mean attendance (i.e. cut-off value) \cite{korutcheva2004advances}. The fluctuation around the mean attendance is known as \textit{volatility}, $\sigma$. Volatility is an inverse measure of the system's performance and hence, the term $\sigma^2/N$ corresponds to an \textit{inverse global efficiency}. When the fluctuations are smaller, that implies that the size of the minority, thus the number of winners, is larger. Hence, smaller volatility corresponds to higher users' satisfaction levels along with better resource utilization. It is known that volatility depends on the ratio $2^m/N$, which is commonly referred to as the \textit{training parameter} or \textit{control parameter} (Let $\alpha = 2^m/N$) \cite{korutcheva2004advances}\cite{Yeung2009}\cite{Johnson1999493}. (An example follows in Section \ref{offloading}, in particular in Fig. \ref{volatility_averaged}.)
 
\item{\em Phase transition}:
Using the variation of the global efficiency w.r.t. $\alpha$ (See also Fig. \ref{volatility_averaged}), it can be seen that the game is divided into two \textit{phases} by the minimum value of $\alpha$ (denoted by $\alpha^*$), namely \textit{crowded phase} and \textit{uncrowded phase}. MG is said to be in the crowded phase when $\alpha < \alpha^*$. This is because, for smaller $m$, the number of strategies, $2^{2^m}$, is quite smaller compared to the number of agents $N$, thus many agents could be using the same strategy, leading them to make the same decision. This then creates a \textit{herding effect}, causing the MG to enter the crowded phase. Once $\alpha > \alpha^*$, the $m$ values are large enough to make the strategy space larger than the number of agents $N$, so that the probability of any two agents using identical strategies diminish, thus making MG enter the uncrowded phase. Note that, $\alpha^*$ corresponds to the minimum volatility indicating the system's ability to self-organize into a state where the number of satisfied agents and the resource utilization are maximized. Moreover, it is shown that the performance of MG surpasses that of the random choice game (where all agents choose each action with a probability $= 0.5$) for a certain range of $\alpha$ values. This is referred to as the \textit{better than random regime} \cite{korutcheva2004advances}\cite{Yeung2009}\cite{Marsili2000522}\cite{Challet:2014:MGI:2616207}.

\item{\em Predictability}:
This is an important physical property of MG. It measures the information content in the previous set of \textit{attendance} values, that is available to agents. Predictability is denoted by $H$, where $H = 0$ corresponds to the situation in which the game outcome is unpredictable. Moreover, the predictability is the parameter that characterizes the two phases in the MG. In MG, $H = 0$ for $\alpha < \alpha^*$ and $H \neq 0$ for $\alpha > \alpha^*$. This implies that during the crowded phase, the game outcome is unpredictable and when MG enters uncrowded phase, the game outcome becomes more predictable \cite{Yeung2009}.


\end{itemize}

\subsection{Equilibrium Notions for an MG}
\label{equilibrium}
In this section, we provide a brief introduction to the notion of equilibria of MG. The reader is encouraged to look further (e.g. \cite{Marsili2000522}, \cite{Challet:2014:MGI:2616207}) for an in-depth tutorial.

Assuming the number of agents is an odd number equal to  $N$, an MG is in an equilibrium if each of the two alternatives is selected by $(N-1)/2$ and $(N+1)/2$ agents. Then, no agent would gain by unilaterally deviating from its state since, if any of the agents in majority group does so, the groups would switch thus the state of the deviated agent would not improve. In an MG stage game, three types of Nash equilibria (NE) are applicable. Note that the NE corresponds to the local minima of volatility values \cite{Challet:2014:MGI:2616207}.

\begin{itemize}  
\item{\em Pure strategy Nash equilibria}: If there are $N$ agents playing the MG and $(N-1)/2$ of them choose to select one alternative with probability $= 1$ while the other $(N+1)/2$ agents select the other alternative with probability $= 1$, system is said to be in a pure strategy NE. There are $\binom{N}{\frac{N-1}{2}} + \binom{N}{\frac{N+1}{2}}$ number of such NEs that exist. These NEs are considered the \textit{globally optimal} states \cite{Marsili2000522}. 
  
\item{\em Symmetric mixed strategy Nash equilibria}: There exists only a single symmetric mixed strategy NE to the MG. It corresponds to the so called \textit{random choice game} where, all agents choose between each of the two alternatives with a probability of $0.5$ \cite{Marsili2000522}. 
\item{\textit{Asymmetric mixed strategy Nash equilibria}:} If $(N-1)/2$ agents select one alternative with probability $= 1$, another $(N-1)/2$ agents select the other alternative with probability $= 1$ and the remaining agent selects an alternative with an arbitrary mixed probability, the MG stage game is said to be in an asymmetric mixed strategy NE. There can be an infinite number of such NEs \cite{Marsili2000522}. 
\end{itemize}

\subsection{Solution Approaches}

Both qualitative and analytical approaches have been studied in the literature to solve an MG. The qualitative approach investigates how the volatility of the system varies with respect to the brain size and the population size. Moreover, it interprets the phase transitions of MG from the crowded phase to the uncrowded phase along with the volatility variation. Hence, this approach is also referred to as \textit{crowd-anticrowd theory} in the literature \cite{Yeung2009}. A brief overview of the phase transition in MG in relation to the variation of volatility is given in Section \ref{MGproperties}. The interested reader can find more rigorous explanations in \cite{Challet:2014:MGI:2616207}.

The analytical solution of MG obtains its statistical characterizations of the stationary states and the NE. In MG, the stationary states correspond to the minima of \textit{predictability} whereas the NE of the MG correspond to the minima of \textit{volatility}. In order to derive the analytical solutions, numerous mathematical techniques are used, including reinforcement learning, replicator dynamics, and tools from statistical physics of disordered systems (e.g. Hamiltonian, replica method, spin glass model, Ising model). Reference \cite{Marsili2000522} includes a rigorous analysis on the solution of MG where aforementioned techniques are employed to derive the solution. Therein, complete statistical characterizations of the stationary state of MG are realized. First, the authors use \textit{multi-population replicator dynamics} technique to obtain some evolutionarily stable NEs of the stage game. Then, a generalized version of the repeated MG model is analyzed where \textit{exponential learning} is used by the agents to adapt their strategies. 

For this scenario, analysis is done considering two different types of agents, namely \textit{naive}\footnote{In seminal MG, the agents are naive, since they only know the pay-off received by the played strategies.} and \textit{sophisticated}.\footnote{Unlike naive agents, sophisticated agents are assumed to have the knowledge of the pay-off they would receive for any strategy that they play (including the strategies that are not played). More details on naive and sophisticated agents can be found in \cite{Marsili2000522}.} In their analysis the authors show that for the repeated MG with naive agents, stationary state is not a NE. Moreover, authors show that for the systems with sophisticated agents and exponential learning, the system converges to a NE.

\subsection{Variants of the Minority Game}
In this section, we briefly introduce few different variations of MG beyond the basic form described before. A comprehensive discussions can be found in \cite{Yeung2009}, \cite{Johnson1999493} and \cite{Mertikopoulos:2007:SGS:1345263.1345272}, among many others. 
\begin{enumerate}

\item \textit{MG with arbitrary cut-off}: A generalized version of the basic MG, referred to as \textit{MG with arbitrary cut-offs} is introduced in \cite{Johnson1999493}. In such games, the minority rule is defined at an arbitrary cut-off value ($\phi$) rather than the $50\%$ cut-off used in the seminal MG. In \cite{Johnson1999493}, authors show the behavioral change of MG when the cut-off value is varied. In brief, it is shown that the attendance values fluctuate around the new cut-off, exhibiting the adaptation of the population. Furthermore, the analysis shows that when the cut-off $\phi$ is decreased below $N/2$, the brain size yielding the minimum volatility is also decreased. This variant is particularly useful to model some resource allocation problems where the \textit{comfort value} (also known as the cut-off) of the capacity of a particular resource is a value other than $50\%$. The MG model for the problem of computation offloading  presented in Section \ref{offloading} in this article counts as an example.

\item \textit{Multiple-choice MG (Simplex game)}: This variant is introduced in \cite{Mertikopoulos:2007:SGS:1345263.1345272} as a direct generalization of the basic MG where every agent might select among $K$ different choices ($K>2$). Thus, a simplex game is defined by the set of $N$ players, the set of $K$ choices and the history winning actions ($m$-bit long). Similar to the seminal MG, a strategy is a mapping of the history data to one of the $K$ choices, and each player is given a set of $S$ strategies. 
Moreover, the strategy space of the simplex game is associated with probability values $p_{is}$, which indicates the satisfaction of the $i^{th}$ agent with her $s^{th}$ strategy. Each player's choice is referred to as a \textit{bid} and a quantity called \textit{aggregate bid} is defined as the sum of the choices of all agents. Similar to the \textit{attendance} property in the basic MG, the aggregate bid contains the information about the number of agents that select a given choice and determines the pay-off that each user receives. 
As the game is played iteratively, after scoring the strategies, the probability values ($p_{is}$) are updated using the \textit{exponential learning}\footnote{In exponential learning, the probability values are modified by following a \textit{logit model}-like formula. This formula contains exponential functions of strategy score and the agent's \textit{learning rate}, which is a numerical constant which might differ for each agent \cite{Marsili2000522}.} method.
Thus, although the players start out naive, they become sophisticated as the game evolves. Therefore, unlike basic MG, the simplex game exhibits evolutionary behavior. In \cite{Mertikopoulos:2007:SGS:1345263.1345272}, it is shown that compared to playing an MG with few options, in a game with a large action set, the overall system performance improves, resulting in higher resource utilization.

\item \textit{Evolutionary MG (Genetic model)}: In this version of MG, unlike the basic case, all users apply a single strategy. Each agent $i$ chooses the action predicted by the strategy with some probability $p_i$ referred to as the agent's \textit{gene value}. Each agent selects the opposite action with probability $1-p_i$. At each play, $+1$ (or $-1$) point is assigned to each agent in the minority (or majority). As the game evolves, if the accumulated score falls below a certain threshold, a new gene value is drawn (known as mutation of gene value) \cite{Yeung2009}. 

\item \textit{Grand canonical MG (GCMG)}: In this type of MG, the number of players who participate in the game can vary since the players have the freedom of being \textit{active} or \textit{inactive} at any round of the game. More precisely, in a GCMG, agents would score their strategies as usual and if the highest strategy score is below a certain threshold, agents would abstain from playing the game for that round of play. Any inactive agent re-enters the game when participation becomes profitable. Consequently, the attendance is calculated based on active players only \cite{Yeung2009}. 

\end{enumerate}

\section{MG Models in Communication Networks: State-of-the-Art and Future Potential Applications}
\label{state_of_art}

In this section, we first provide a brief summary of the state-of-the-art, where MG is applied to solve the problems that arise in communication networks. Future research directions and open problems are discussed as well.

\subsection{MG Models in Communication Networks: State-of-the-Art}

\subsubsection{Interference management} 
In \cite{5671703}, the authors investigated distributed interference management in Cognitive Radio (CR) networks using a novel MG-driven approach. They propose a decentralized transmission control policy for secondary users, who share the spectrum with primary users thus causing interference. In this work, secondary users play an MG, selecting between two options, namely \textit{transmit} or \textit{not transmit}. The winning group is determined based on the interference experienced by the primary user. For instance, if the majority transmits, the interference power measured at the primary receiver exceeds the threshold so that the minority who does not transmit become the winners and vice versa, ensuring that the minority always wins. At each round of play, the primary receiver announces the winning group through sending a control bit to secondary transmitters.
  

\subsubsection{Wireless resource allocation and opportunistic spectrum access}
An example of wireless channel allocation using MG can be found in \cite{6884175} where an MG-based mechanism for energy-efficient spectrum sensing in cognitive radio networks (CRNs) is presented. The authors emphasize how the MG, due to its self-organizing nature, befits to model such problems to achieve cooperation and coordination gain without causing a large signaling cost in a CRN. In the applied MG model, the agents are the secondary users, who choose between \textit{sensing} or \textit{not sensing}, in the process of detecting an idle channel. Two different distributed learning algorithms are then developed that are applied by the agents to converge into equilibrium states characterized by pure and mixed strategy NE. In \cite{4604741}, a multiple-choice MG (simplex game) is used to model the resource allocation problem in heterogeneous networks, where a large number of non-cooperative users compete for limited radio resources. The existence of correlated equilibrium was proved. Moreover, the authors compared the equilibria with the optimal states using the concept of the price of anarchy. 

\subsubsection{Coordination in delay tolerant networks}
In \cite{6576421}, an MG-based model was applied to coordinate the relay activation in delay tolerant networks in order to guarantee an efficient resource consumption. In the MG model, relays act as the players who decide to \textit{transmit} (participate in relaying) or \textit{not to transmit} (not to participate in relaying). In their work, the authors developed a stochastic learning algorithm that converges to a desired equilibrium solution. 


\subsection{Potential Future Applications and Open Problems}

As discussed in the previous section, most of the existing work is focused on the application of MG models in cognitive radio networks. The applications of MG for 5G SCNs remain however unexplored. In what follows, we discuss some possible applications as well as open theoretical issues. 

\subsubsection{Computation offloading in small cell networks}
 With the emergence of new mobile applications, it has become common for small user devices to have computationally-intensive tasks, such as image processing to perform.  However, due to limited computational capability and limited battery capacity, user devices are not always capable of performing the desired task, or doing so might become  inefficient. This gives rise to the idea of offloading such tasks to a remote server (e.g. the cloud), which typically has much higher computational capability than local devices. This idea is referred to as \textit{computational offloading}. Computational offloading is expected to save the energy cost spent for local execution, thereby saving the battery life of end user devices.

Despite great benefits, computation offloading also imposes certain challenges. These include large communication costs in terms of energy and latency, caused by the long distance between users and cloud servers, which are typically located outside the local network. Moreover, especially in dense SCNs, excessive back-haul traffic might arise as a result of the large number of offloading requests being sent to cloud.
Thus, in such cases, computational offloading to nearby SBSs (known as \textit{mobile-edge offloading}) can be a better alternative for users. On the other hand, utilizing the SBS resources for computational offloading is an efficient way to make use of the idle resources located in the widely available SBSs, especially in dense networks \cite{s16070974}.

The applicability of MG to study computational offloading problem is well-justified: It is essentially a dynamic resource allocation problem where users, non-cooperatively and selfishly, try to utilize the limited computational resources located in the SBS.  Moreover, users cannot communicate with each other or observe each others’ actions. While any centralized approach to solve such problem might be very inefficient,  an approach based on MG model is distributed and of low cost. Besides, in an MG model, all players eventually exhibit cooperative behavior as a collection, despite being selfish individually. 

In Section~\ref{offloading}, we will present a novel MG-based model to address the computational offloading problem in dense SCNs. 
 
\subsubsection{Transmission mode selection} 
Device-to-Device (D2D) communication is considered as a building block in 5G networks. In D2D network underlaying an SCN, users have two modes of transmission: (i) direct communication without using the core network infrastructure, (ii) communication with the aid of the base station as in regular cellular networks. Clearly, the transmission mode selection problem can be modeled as an MG, where users are modeled as agents and the two options correspond to transmission via \textit{D2D mode} or \textit{cellular mode}. Given limited resource, the reward of each mode (for instance the throughput) then depends on the number of users selecting that mode. Thus, the cut-off value that defines the minority depends on factors such as number of interferers, number of available direct channels, etc.

\subsubsection{Multiple-choice games} It is clear that the current state-of-the-art mostly use the basic MG, which limits the agents to select between only two alternatives. Nonetheless, in many practical resource allocation problems, the decision is made among multiple choices. Examples include the channel selection or computation offloading problems where a channel/computational server is selected from many potential options.

\subsubsection{Evolutionary variations of MG} 
In a basic MG, the agents' selected set of strategies do not evolve. In other words, they stay fixed through out the iterations and are only scored in each play so that the agents can learn the best strategy for them. Hence, the basic iterated MG cannot be classified as an evolutionary game, making it inapplicable to model the problems where users should have the capability of altering their given strategies. On the other hand, complex dynamic systems require the agents to not only learn the best strategy but also to \textit{adjust} the strategies as the game advances in a dynamic manner. To accomplish this, the modified versions of the seminal MG such as evolutionary MG (EMG) can be used. Other options include MG models that use learning methods such as \textit{exponential learning} to adjust the strategies. The current state-of-the-art consists of very few applications of such evolutionary variations of the MG, thus it can be noted as a potential research direction. 



\subsubsection{MGs for players with heterogeneity}
 In a practical point of view, it is very likely for the SCN users to be heterogeneous and to have diverse QoS requirements (e.g. in terms of delay, rate and energy efficiency). However, in the basic MG, every player is assumed to have similar capabilities and uses identical information and learning methods. Thus, generalization of the basic MG model to include such heterogeneities among users can be considered as a potential line of research. 

\section{Computation Offloading in Small Cell Networks: When Minority Wins}
\label{offloading}

\subsection{System Model and Assumptions}
We use a minority game with an arbitrary cut-off. Consider an SBS serving $N$ number of homogeneous (with respect to both  computational capability and the task potentially to be offloaded) users. Each computational offloading period $t$ is considered to be a round of play of the MG. All users participate as the players of the game and individually decide whether they offload the task to the local SBS or they execute it locally using their own resources. Users  select one of these two options simultaneously within each offloading period and they have no information about other users' actions. Note that users naturally prefer to offload to the local SBS rather than executing the tasks locally, provided that the local SBS does not become crowded with offloading requests. The reason is as follows. Analogous to the original bar problem \cite{RePEc:aea:aecrev:v:84:y:1994:i:2:p:406-11} where the customers naturally like to go to the bar than staying home if the bar is uncrowded, we assume that by offloading to an uncrowded SBS, users can experience lower latency. The SBS supports all of the computation offloading requests it receives, by completing all tasks in a TDMA manner. (This can be done using virtual parallel processing, where the time slot given to each task is small enough to assume that all tasks are performed simultaneously.) Therefore, if the number of offloading requests exceeds a certain threshold, the latency experienced by users might increase, making local computation to become the preferred option. Note that for the sake of simplicity, we assume the amount of energy required for the local computation is approximately equal to the amount of transmission energy required for the offloading. Thus we omit the energy parameter in this simplified model.

The problem described above is a distributed resource allocation problem where we analyze how to optimally utilize the computational resources of the local SBS while the latency remains below a specific threshold. As conventional, we assume that the local SBS has a fixed computational capability. In each round, users have the two options of either offloading or locally computing. The number of offloading requests that the SBS can handle is an arbitrary cut-off value denoted by $\phi$. Clearly, this offloading threshold ($\phi$) counts as the minority rule of the game. Note that $\phi$ remains unknown to the users throughout the game. For every user, $L_{th}$ is the maximum tolerable latency, which is experienced if the computation is performed locally. Then the cut-off value $\phi$ is defined such that, when the number of offloading users approaches $\phi$, the latency for offloading users reaches the threshold, $L_{th}$. Thus, for a user to benefit from offloading, the number of offloading users should not exceed the limit of $\phi$. As a result, being in the population minority (defined by $\phi$) is always desired. After each round of play, one of the  outcomes mentioned below would occur.

\begin{itemize}
\item \textit{\textbf{If minority chooses to offload and majority chooses to compute locally}}:
In this case, the minority receives a reward since offloading yields lower latency than local computation since the local SBS is uncrowded. 
\item \textit{\textbf{If minority chooses to locally compute and majority chooses to offload}}:
In this case, the number of offloading requests exceeds the threshold $\phi$ so that the SBS becomes too crowded. Consequently, the latency for the offloading users would exceed the allowable latency threshold. Thus minority wins and receives a reward.
\end{itemize}

\subsection{MG Model}

\subsubsection{Attendance}
Conventionally in MG, the winning choice is announced to all users after each round of play, so that users take advantage of this information to score their strategies. Accordingly, in our model, after each round of play, the SBS broadcasts the winning group by sending an one-bit control information $b(t)$ defined as
\begin{align}
    b(t)= 
\begin{cases}
    1,& \text{if } n(t)<\phi\\
    0,& \text{if } n(t)\geq \phi
\end{cases}
\end{align}
where $n(t) = $ number of offloading users at offloading period $t$. 

In accordance with the MG terminology, we refer to this $n(t)$ as the \textit{attendance}. Given the control information, users evaluate their strategies to improve decision making in the next round of play.


\subsubsection{Reward}
The reward of each winning user is defined based on the computation latency experienced by the user. Note that the transmission delays and propagation delays are considered negligible compared to computation latency. Thus the reward depends on the number of other users who select the same option, be it offloading or local computation.
\begin{itemize}
\item $L(t)$ = Latency experienced by an offloading user at offloading period $t$,
\item $C_{b}$ = Computation capability of SBS (in number of CPU cycles per unit time), 
\item $C_{u}$ = Computation capability of local user device (in number of CPU cycles per unit time),
\item $M$ = Number of CPU cycles required to complete the task.
\end{itemize}

Thus the latency experienced by an offloading user yields $L(t)$ = $n(t) \cdot M/C_{b}$, and the latency experienced by as locally computing user is given as $L_{th}$ = $M/C_{u} $. 
For $L(t)$ = $L_{th}$, $n(t)$ = $\phi$, hence $\phi = C_{b}/C_{u}$. If $L(t) \geq L_{th}$ (equivalent to $n(t) \geq \phi $), offloading users (majority) lose and locally computing users (minority) win. In contrast, when $L < L_{th}$ (i.e. $n(t) < \phi$), offloading users (minority) win and locally computing users (majority) lose. Let $U_o(t)$ and $U_l(t)$ be the utility each user, respectively receives in case of offloading and local computing. Thus we have

\begin{align}
    U_o(t) = b(t) =  
\begin{cases}
    1,& \text{if } n(t)<\phi\\
    0,& \text{if } n(t)\geq \phi
\end{cases}
\end{align}

and
\begin{align}
		U_l(t) =
\begin{cases}
    0,& \text{if } n(t)<\phi\\
    1,& \text{if } n(t)\geq \phi.
\end{cases}
\end{align}

\subsubsection{Distributed Learning Algorithm}
 
To solve the designed MG, we use the reinforcement technique \cite{RePEc:eee:phsmap:v:246:y:1997:i:3:p:407-418}\cite{Yeung2009}. For the comparison purposes, MG-based method is compared with a random selection scenario. 

\begin{algorithm}
\label{Algo}
\caption{Distributed learning algorithm to solve offloading MG \cite{RePEc:eee:phsmap:v:246:y:1997:i:3:p:407-418}}
\begin{algorithmic}
\STATE \textbf{\textit{Initialization}}: Each user $i$ randomly draws $S$ strategies.
\FOR{$t=2:T $}
\STATE Each user $i$ selects action $a_i(t)$ predicted by the best strategy.
\STATE SBS broadcasts the control information $b(t)$.
\FOR{$s=1:S$}
\STATE Each user $i$ updates the score of the strategy $s, V_{i,s}$.
\IF{prediction of $s=b(t)$} 
\STATE $V_{i,s}(t+1) = V_{i,s}(t) + 1$, 
\ELSE
\STATE $V_{i,s}(t+1) = V_{i,s}(t)$.
\ENDIF
\ENDFOR
\STATE Each user $i$ selects best strategy $s_i(t)$, defined as $\argmax\limits_{s \in \mathcal{S}} V_{i,s}(t)$
\STATE $t \gets t+1$
\ENDFOR
\end{algorithmic}
\end{algorithm}

It is shown in \cite{Marsili2000522} that using this basic MG model results in naive behavior of agents since they do not account for their \textit{market impact}\footnote{In this model, agents respond to the aggregate action of \textit{all} agents in the previous round, which also contains their \textit{own} actions. Thus the naive agents do not account for their own actions, which is referred to as \textit{market impact} \cite{Marsili2000522}.}, which makes the system unable to attain NE.
In this work, we simply use the self-organizing capability of basic MG around the cut-off, even with its naive agents. Future work can explore more generalized versions of MG models that attain NE
 with sophisticated agents who account for market impact.

\subsection{Simulation Results and Discussion}
For numerical analysis, we consider an SBS serving $N=31$ users. The task that the users have to perform is assumed to require $M = 10$ Megacycles of CPU cycles. The CPU capacity of each user device is $C_{u}=0.5$ GHz. Moreover, the SBS allocates $C_{b}=10$ GHz of CPU capacity to serve users' offloading requests. Hence the system's cut-off value becomes $\phi=C_{b}/C_{u}=20$. Thus, if the attendance is less than $20$, the offloading users win, and vice versa. We simulate the system for different brain sizes ($m$) to observe the system behavior. For each $m$ value, $32$ runs are carried out, where each time, users randomly draw a new set of strategies ($S=2$). In each of the $32$ runs, $T=10000$ offloading periods (rounds of MG) are executed. For comparison, we also implement the random choice game where users select one of the two actions (to offload or not) with equal probabilities at each round of the game.

\begin{figure}[!htb]
  \centering
	\includegraphics[width=0.44 \textwidth]{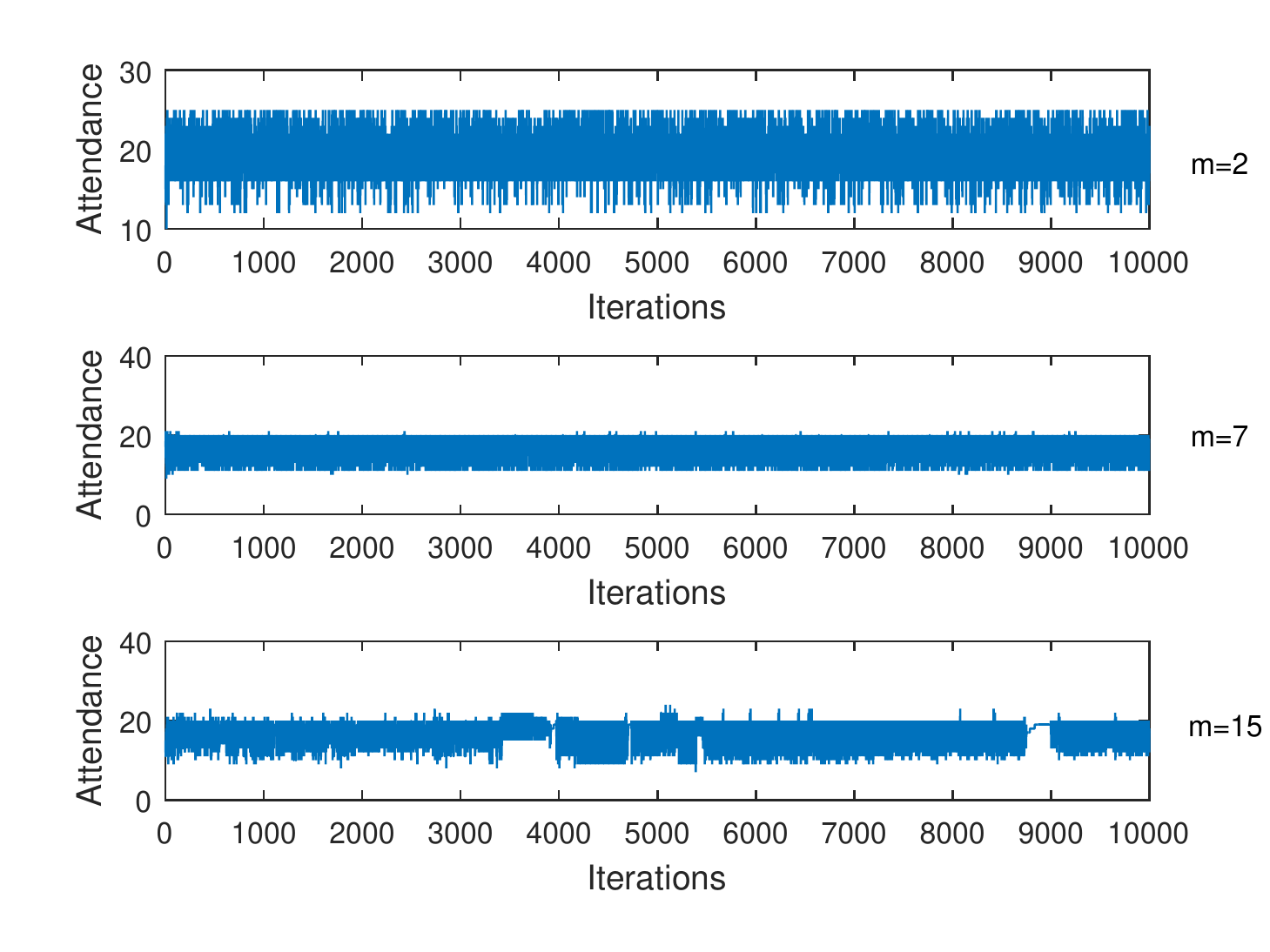}
   \caption{Time evolution of attendance}
	\label{time_evolution_of_attendance}
\end{figure}

The variation of {\em attendance} over time for different brain sizes ($m$) are shown in Fig. \ref{time_evolution_of_attendance}. From the figure it is clear that the number of offloading users always fluctuate near the cut-off value, when users play an MG. It implies that the system self-organizes into a state where the number of users who experience a latency less than the threshold is near its maximum value, thereby maintaining the optimal SBS utilization. This is interesting since the users are not given any prior information about the exact cut-off value of the system. As mentioned earlier, fluctuation of the attendance (i.e., the standard deviation) is known as the volatility in MG literature. It is clear that for different $m$ values, the amount of fluctuation differs as explained in Fig. \ref{volatility_averaged} (see below). Note that the fluctuations correspond to the amount of cooperation in the MG. They provide a measure for the number of users who could have offloaded if the attendance is lower than the cut-off or for the number of agents who could have not selected offloading option, if the attendance exceeds the cut-off. 
From the attendance figures, it can be seen that, even though the cut-off value is not advertised to the agents, the population adapts to the cut-off value of the system. 
The reason for this behavior is the agents' adaptation to the environment they collectively create.  

Fig. \ref{volatility_averaged} shows that the variation of the standard deviation ($\sigma/N$) of the number of offloading users over different $m$ values follows the expected MG behavior \cite{korutcheva2004advances}\cite{Yeung2009}\cite{Johnson1999493}, described in the following. As discussed, the volatility corresponds to the fluctuations of the attendance and serves as an established measure for the system performance. Lower volatility values mean that the fluctuations around the cut-off decrease. This corresponds to the size of the minority being larger, resulting in a larger number of winners, thus a better performance. Accordingly, lower volatility corresponds to better resource utilization and higher user satisfaction. From Fig.~\ref{volatility_averaged}, for almost all values of $m$, volatility is lower than that of the random choice game. Hence one can conclude that the resource utilization is improved when MG-based offloading method is used. 
This shows the self-organizing nature of the MG, where agents coordinate to reduce the fluctuations in the absence of any communication or information other than the history data. It can also be seen that a minimum of the average volatility occurs at $m=3$, where the phase transition from the \textit{crowded phase} to the \textit{uncrowded phase} occurs. 

\begin{figure}[!htb]
  \centering
	\includegraphics[width=0.44 \textwidth]{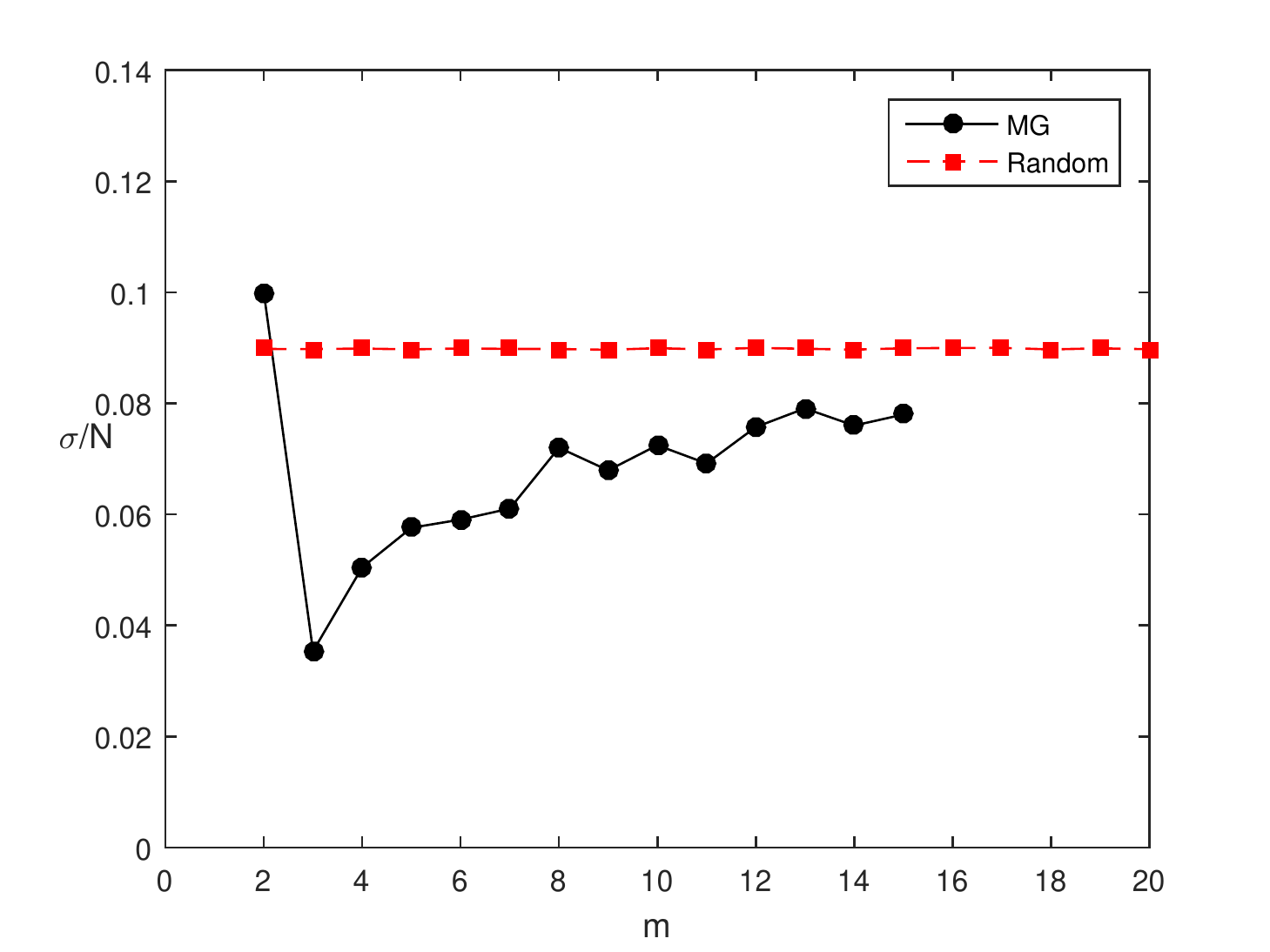}
   \caption{Variation of volatility  with $m$}
	\label{volatility_averaged}
\end{figure}

\begin{figure}[!htb]
  \centering
    \includegraphics[width=0.44\textwidth]{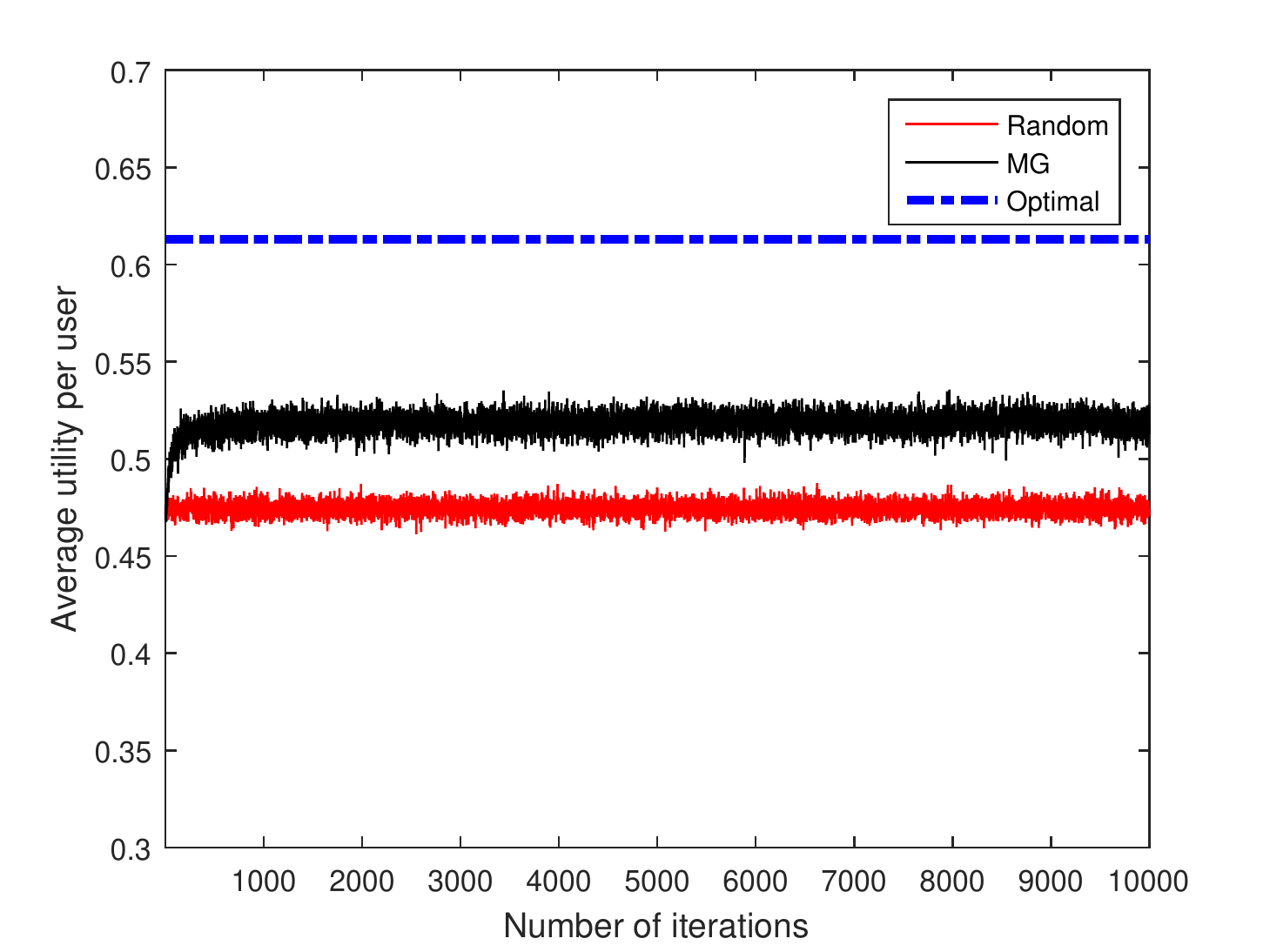}
		\caption{Average utility received by users}
		\label{utility}
\end{figure}
\begin{figure}[!htb]
  \centering
    \includegraphics[width=0.44\textwidth]{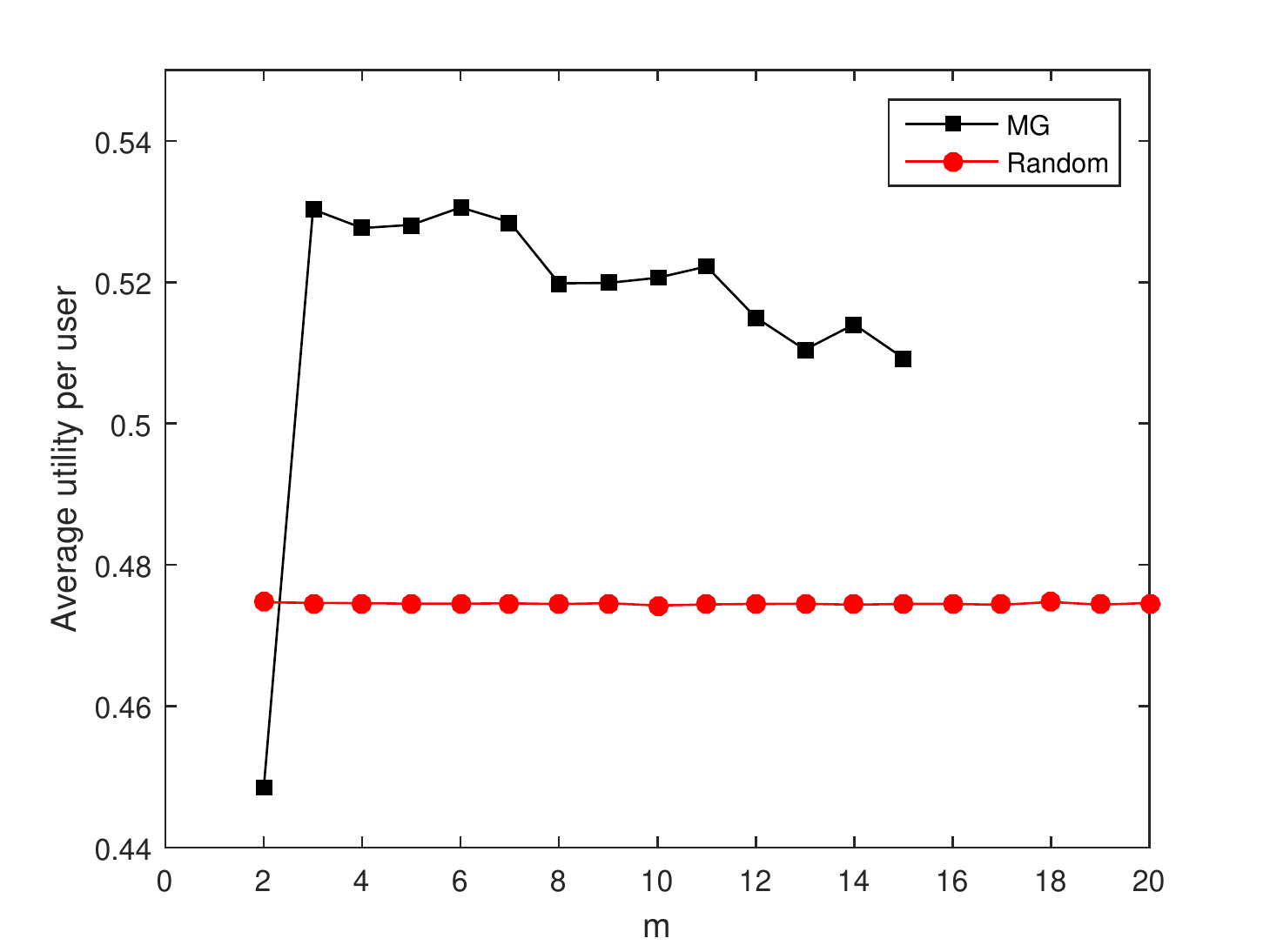}
		\caption{Average utility vs. $m$}
		\label{utility_vs_m}
\end{figure}

To investigate the improvement in the latency experienced by the users, average utility is shown in Fig. \ref{utility}. It is clear that for MG, the utility achieved by individual users is better than that of the random choice game. However, it can be seen that the average utility received by a user who applies MG-based offloading is still lower than that of an optimal situation, where the number of offloading users is always equal to $\phi-1$ (here $19$). 
This is the price of the lack of coordination between agents and the use of minimal external information. Fig. \ref{utility_vs_m} depicts the influence of the brain size on the average utility achieved per user. As expected, for larger volatility values, the achieved utility is substantially smaller. Roughly speaking, Fig. \ref{utility_vs_m} is approximately an inverse of the volatility figure (Fig. \ref{volatility_averaged}). Thus, once more we come to the conclusion that the volatility is indeed an inverse performance measure for the MG-based system.

\begin{figure}[!htb]
  \centering
    \includegraphics[width=0.44\textwidth]{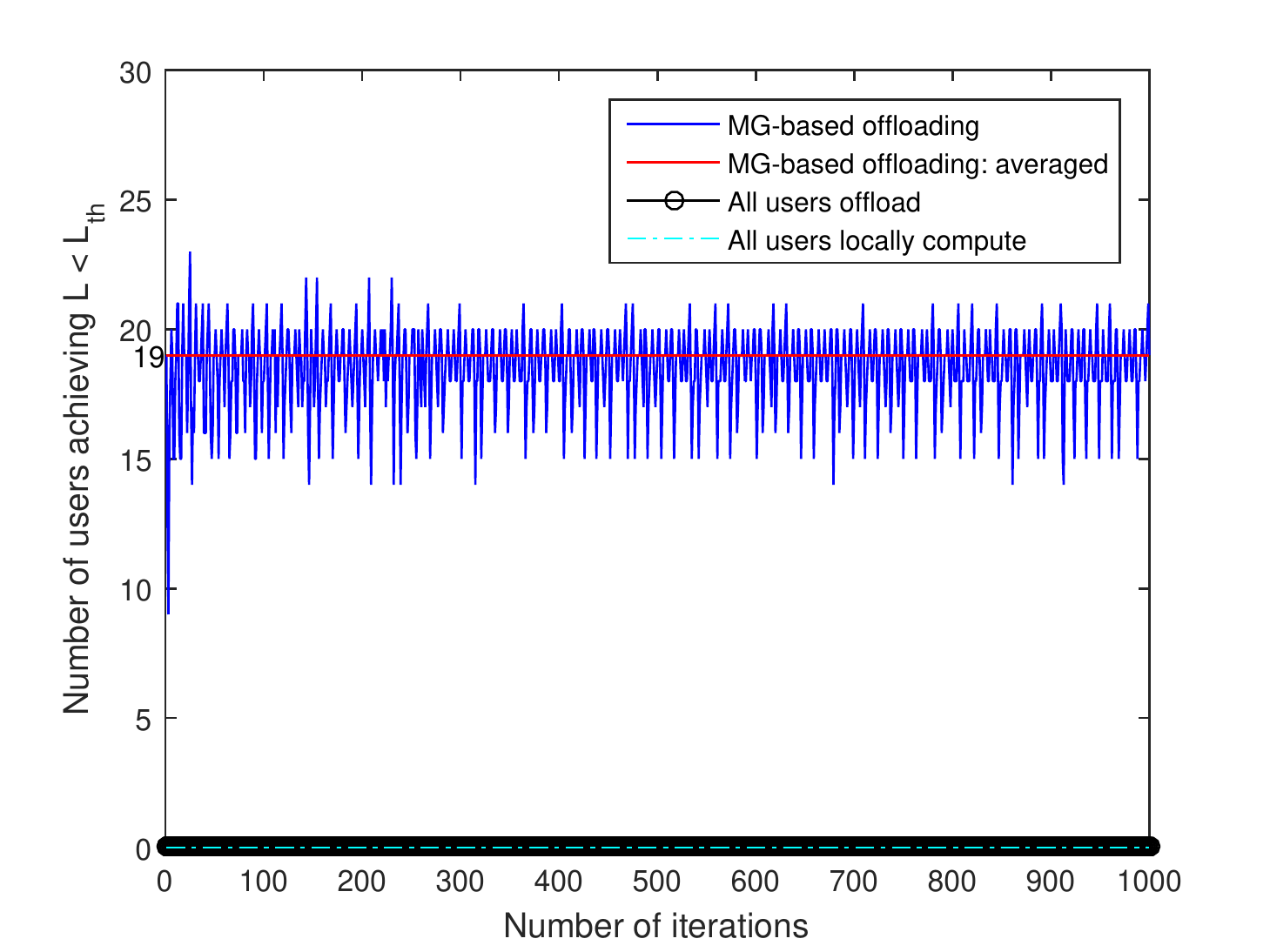}
		\caption{Number of users achieving below-threshold latency}
		\label{comparison}
\end{figure}

In Fig. \ref{comparison}, we illustrate the benefit of MG-based offloading. When the MG-based offloading mechanism is used, the number of users who experience latency below the threshold fluctuates near its maximum value  ($\phi-1=19$, in this example), compared to the two general cases where all users simply offload or compute the task locally. In the last two cases, none of the users is able to achieve a latency below the threshold. Using our defined model, if all users offload, the latency experienced by each user yields $31$ milliseconds. Similarly, if all users choose to compute locally, the latency is $20$ milliseconds. Since the threshold latency is $20$ milliseconds, it is clear that none of the above two methods would allow the users to experience latency values \textit{below} the threshold. Consequently, using an MG-based approach results in achieving some latency below the threshold for a larger number of users, thereby utilizing the available SBS resources in a productive manner.




\section{Conclusion}
We have presented the basics of minority game models and their applications in communication networks. Also, future potential applications and open research issues have been outlined. As a potential application of minority games in 5G small cell networks, we have investigated the distributed mobile computation offloading problem and some preliminary results have been presented.
In mobile-edge computation offloading, users typically have several resources to offload to; these include local device, SBS, MBS, D2D offloading or the cloud \cite{s16070974}. To model such cases, multi-option MG (simplex game) can be used. Also, in practice, users can have a variety of devices with varying computational and battery capacities (e.g., smart phones, tablets, laptops); thus MG models that incorporate different user types can be employed to model such scenarios. Apart from the conventional method, different learning techniques such as biased \cite{Yip2003318} and adaptive \cite{4274903} strategies can be adopted to achieve better performance. 

\bibliographystyle{IEEEtran}

\bibliography{magazine_ref}

\end{document}